\documentclass[preprint,superscriptaddress, showpacs, floatfix, nobalancelastpage]{revtex4}
\usepackage{amsfonts}
\usepackage{amssymb}
\usepackage[dvips]{graphicx}
\usepackage{color}
\usepackage{amsmath}
\usepackage[bookmarksnumbered, bookmarks, breaklinks, linktocpage]{hyperref}
\begin{document}

\title{Magneto-elastic Effects in $\text{Tb}_{3}\text{Ga}_{5}\text{O}_{12}$}
\author{U. L\"ow}
\affiliation{Theoretische Physik II, Technische Universit\"at Dortmund,
44227 Dortmund, Germany}
\author{S. Zherlitsyn}
\affiliation{Hochfeld-Magnetlabor Dresden, Helmholtz-Zentrum Dresden-Rossendorf, D-01314 Dresden, Germany}
\author{K. Araki}
\affiliation{Institute for Solid State Physics, University of Tokyo, Kashiwa, Chiba 277-8581, Japan}
\author{M. Akatsu}
\affiliation{Graduate School of Science and Technology, Niigata University, Niigata 950-2181, Japan}
\author{Y. Nemoto}
\affiliation{Graduate School of Science and Technology, Niigata University, Niigata 950-2181, Japan}
\author{T. Goto}
\affiliation{Graduate School of Science and Technology, Niigata University, Niigata 950-2181, Japan}
\author{U. Zeitler}
\affiliation{High Field Magnet Laboratory and Institute for Molecules and Materials, Radboud University Nijmegen, Toernooiveld 7, 6525 ED Nijmegen, The Netherlands}
\author{B. L\"uthi}
\affiliation{Physikalisches Institut, Johann Wolfgang Goethe Univerist\"at Frankfurt, D-60438 Frankfurt (M), Germany}
\date{\today}

\begin{abstract}
We report new results for the elastic constants studied in Faraday and Cotton-Mouton geometry in 
Tb$_3$Ga$_5$O$_{12}$ (TGG), a frustrated magnetic substance with the strong spin-phonon interaction and remarkable crystal-electric-field (CEF) effects. We analyze the data in the framework of CEF theory taking into account the individual surroundings of the six inequivalent Tb$^{3+}$-ion positions. This theory describes both, elastic constants in the magnetic field and as a function of temperature. Moreover we present sound-attenuation data for the acoustic Cotton-Mouton effect in TGG.

72.55.+s, 73.50.Rb, 62.65+k

\end{abstract}


\maketitle

\section{Introduction }
\label{Intro}

Tb$_3$Ga$_5$O$_{12}$ (TGG)  is a dielectric material with the cubic garnet
structure.  The garnet structure materials show a wide spectrum of physical 
properties: interesting magnetic properties are found in  the ferrimagnetic YIG or 
the rare earth series RIG (with R a heavy rare earth element) or the special DAG
(dysprosium aluminum garnet), but also Laser properties such as in RAlG are found.
For an early review see Ref.~\cite{winkler}. 

The garnet material TGG, described
here, was in the center of interest in recent years. Unconventional
experiments were carried out with this substance, e.g. the so-called phonon
Hall effect \cite{Strohm2005,Inyushkin2007} and the acoustic Faraday effect \cite{us1,us2}. Most recently a
detailed study of magnetic properties has been performed including ESR experiments to analyze 
the crystal electric field (CEF) of the Tb$^{3+}$-ion \cite{us3}. Using elastic neutron scattering, an antiferromagnetic (AFM) transition
was observed at $T_{N} = 0.35$ K, which  was  much lower than the Curie-Weiss temperature, $\Theta{_{CW}}$ $\sim$ 8.61 K, evidencing very high level of magnetic frustrations \cite{Kama}.

In the present paper we proceed with the description of the CEF developed in our earlier work \cite{us3} 
and generalize it to include strain phenomena such as the temperature and magnetic field dependence of the elastic constants. Note, that only a simple cubic CEF model was used before to describe the temperature dependence of elastic constants in TGG \cite{araki}. Furthermore, we show magneto-acoustic birefringence data
and fits discussing new experimental results obtained both in the Faraday and Cotton-Mouton geometry.

The ultrasound experiments have been performed on a TGG single crystal oriented for propagating the sound wave with wave
vector $\vec k$ along the [100] direction. The same sample as in Ref.~\cite{us2} has been used in these experiments. The sample length along
the direction of the sound-wave propagation was $L_0$ = 4.005 mm. The sound velocity and attenuation have been measured with a setup as described at great length in Ref.~\cite{BL}. LiNbO$_3$ transducers have been used in these experiments. Fields up to 33 T have been provided by a resistive magnet at the High Field Magnet Laboratory at Radboud University Nijmegen. The magnet has been equipped with a $^{3}$He cryostat.

The paper is organized as follows. In the next chapter we discuss the Tb$_3$Ga$_5$O$_{12}$ (TGG) CEF model and include the quadrupolar 
operators which are necessary to describe the magneto-elastic interaction. Then we show and discuss the temperature and magnetic field dependence of the elastic constants.
We treat specifically the high field behaviour of the $c_{44}$ mode which was measured in the acoustic Faraday
and Cotton-Mouton geometry.  The effect of the low lying quasi doublet on the elastic modes will 
be investigated because of its relevance for the phonon Hall effect. The magneto-elastic coupling constants gained from these discussions can be used 
for the interpretation of the phonon Hall effect. In addition to the Faraday effect \cite{us1,us2} we present also sound-attenuation data for the Cotton-Mouton-Voigt effect.
\begin{figure}
\begin{center}
\includegraphics[width=14cm,angle=0]{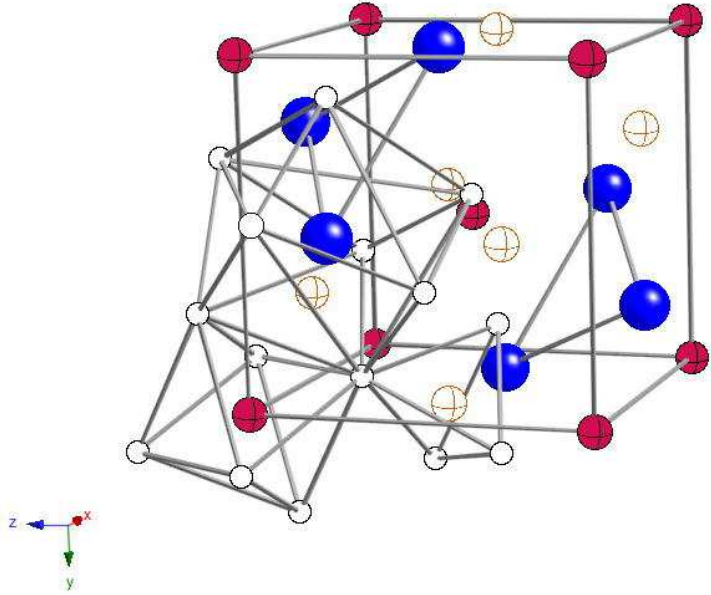}
\end{center}
\caption{(Color on line) Crystal-structure of TGG: Tb-ions blue (large filled spheres),
  Ga-ions red (small filled spheres) and O-ions white (empty spheres). }
\label{fig1}
\end{figure}

\section{The Crystal electric field in TGG}
\label{CEF}

In Fig.~\ref{fig1}  the structure of TGG is shown. The Ga ions are located on cubic corner points whereas the Tb$^{3+}$ ions form corner sharing triangles.
The $\text{Tb}^{3+}$ ions have eight 4$f$-electrons leading with Hund's rule to $S=3$, $L=3$ and $J=6$. Each $\text{Tb}^{3+}$ ion has the same
orthorhombic $D_2$ symmetry in its own local coordinate system and can be described by the crystal field  Hamiltonian, introduced by Guillot et al. \cite{Guillot}:

\begin{eqnarray}
\label{H1}
{H}=&
                                        b_{20}   {\cal O}_{20}
                                       +b_{22}   {\cal O}_{22}
                                       +b_{40}   {\cal O}_{40}
                                       +b_{42}   {\cal O}_{42}
                                       +b_{44}   {\cal O}_{44}\\\nonumber
                                       +& b_{60}   {\cal O}_{60}
                                       +b_{62}   {\cal O}_{62}
                                       +b_{64}   {\cal O}_{64}
                                       +b_{66}   {\cal O}_{66}
+ g \mu_B \vec B \vec J.
\end{eqnarray}
Here $\vec B$ is the magnetic field in the local  coordinate 
system  of a $\text{Tb}^{3+}$ ion.
The ${\cal O}_{ij}$ are the Stevens operators \cite{Hutchings64},
and the $b_{ij}$ are the crystal field parameters
\cite{Guillot}, \cite{Wybourne}.
 The table of the crystal field parameters and the explicit form of the
Stevens operators are given in the Appendix (eqs.~\ref{A1}-\ref{A7}).

In this paper  we use the Hamiltonian with
orthorhombic point symmetry given by eq.~\ref{H1} and take into account the six
inequivalent ion positions in the unit cell \cite{Levitin} to calculate the elastic constants of TGG.
We found it most convenient to rotate the Hamiltonians of eq.~\ref{H1} from the six local coordinate systems which we denote by $l_i$ with
$i=1,\dots 6$ to the laboratory system and then perform all the calculations in 
the laboratory system. In particular for the magneto-elastic coupling this is the best way to proceed, because thus there is no need to transform the
calculated elastic constants back to the laboratory system. For these calculations we used the rotation matrices 
$ R(\alpha_i, \beta_i, \gamma_i)$ where  $\alpha_i, \beta_i, \gamma_i$ 
are the Euler angles. Following the notation of Edmonds \cite{Edmonds} 
the Euler angles of the rotations from the local systems $l_i$ 
to the laboratory system can be easily obtained and are explicitly given in the Appendix (eq.~\ref{table1}).

The rotated Hamiltonians, calculated by use of Mathematica,
have in general complex coefficients.  
In contrast to the original Hamiltonian eq.~\ref{H1} also 
operators  of the type $ {\cal O}_{l,+j} $  and $ {\cal O}_{l,-j} $ appear.    
For these more general cases for which the operators 
are not listed  in \cite{Hutchings64}
we use the form of the operators given by P.A. Lindg\r ard, O. Danielsen  
\cite{Lind}.
As an example, in the Appendix (eq.~\ref{table2}) we list the coefficients $b^{l_1}_{i\pm j}$ 
of the resulting Hamiltonian ${\cal H}^{l_1}$ obtained by rotating 
eq.\ref{H1} with $\vec B=0$ from $l_1$ to the laboratory system. 

\begin{figure} 
\begin{center}
\includegraphics[width=14cm,angle=0]{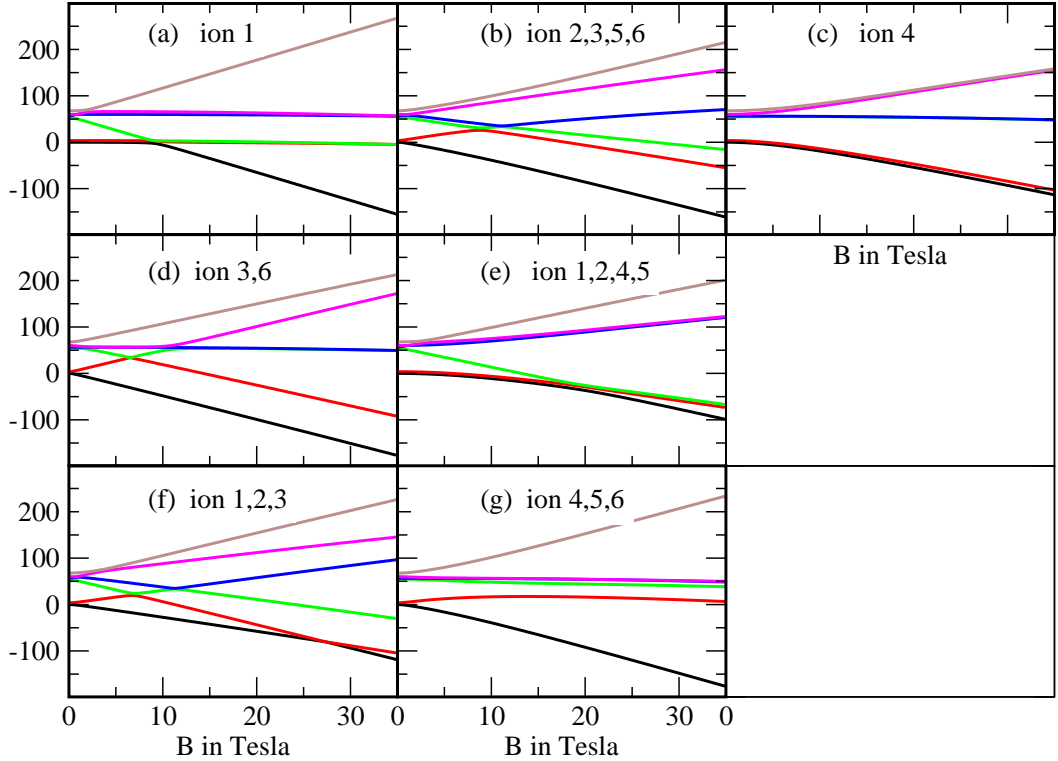}
\end{center}
\caption{(Color on line) Panels a,b,c show the six lowest energy levels for field in $[110]_c$
  direction, panels d,e are the corresponding levels for field in $[100]_c$ direction , and panels f,g 
  for field in $[111]_c$ direction. The corresponding ions are marked in the
  figure, the energies are in Kelvin and the
  magnetic field is in Tesla.} 
\label{fig2}
\end{figure}

In \cite{us3} we calculated the energy levels of the different ions in a magnetic field  
parallel to the cubic $[110]_c$, $[100]_c$ and  $[111]_c$ direction.
Here in Fig.~\ref{fig2} we show the energy levels for the convenience again.

\section{The Magneto-Elastic Coupling}
\label{MEC}

Next we turn our attention to the elastic constants. For an overview of various magneto-elastic couplings see Ref.~\cite{BL}.
We note that, the temperature dependence of the elastic constants 
has been previously described in a cubic approximation \cite{araki}. 
Here we refine the description by incorporating the local surroundings of the 
ions in the framework described in chapter \ref{CEF}. Using this approach we investigate longitudinal and transverse elastic constants
 $c_{11}$, $c' = (c_{11}-c_{12})/2$, $c_{44}$, the bulk modulus $c_B$, and their couplings to the corresponding quadrupole operators.

Profiting from the rotation of the local Hamiltonians we have performed, it suffices 
now to consider the cubic symmetry strains in the laboratory system:
the volume strain  $e_V = e_{xx}+e_{yy}+e_{zz}$
with $c_B=(c_{11}+2c_{12})/3$, the $\Gamma_3$-strain
$e_2=(e_{xx}-e_{yy})/\sqrt{2}$ with 
 $c' = (c_{11}-c_{12})/2$, and the $\Gamma_5$ shear strain $e_{xy}$.

The magneto-elastic Hamiltonian in the laboratory coordinate system reads 
\begin{eqnarray}
\label{H3}
H_{me}   = g(\Gamma_3) e_{2} {\cal O}_{22} + g(\Gamma_5) e_{xy} {\cal O}_{xy},
\end{eqnarray}
where $g(\Gamma_3)$ and $g(\Gamma_5)$ are coupling constants
determined from the experiment and
${\cal  O}_{xy}=\frac{1}{2}(J_x J_y +J_x J_y )$  
and ${\cal  O}_{22}= (J_x^2  -J_y^2)$  
are quadrupole operators.
Here we  calculate
using magneto-elastic Hamiltonian  $H_{me}$  of eq.~\ref{H3}
the elastic constants $ (c_{11}-c_{12})/2$
and $c_{44}$. The case of $c_B$ will be mentioned later.

Since in an ultrasonic wave the strains are small  one can use perturbation 
theory to calculate the strain dependence of the CEF energy levels 
$E_n(\epsilon_\Gamma)$ and of the free energy (see ref. \cite{BL}).
The second order terms in this expansion are the 
the strain susceptibilities $\chi(T)$ 
defined  (in analogy to the magnetic
susceptibility)  as the response of a structural order parameter 
${<\cal O}>$ to an applied strain $e$: 
\begin{eqnarray}
\label{sus_BL}
\chi(T)=\frac{1}{N g^2} \Delta c  = \frac{1}{N g^2}\frac{\partial^2 F}{\partial e^2} \Big|_{e=0} 
=&\frac{1}{Z^2}\frac{1}{k_B T} \left( \sum_n \langle n| {\cal O} | n\rangle
\exp{(\frac{-E_n}{k_BT})}\right)^2\\ \nonumber 
&-\frac{1}{Z} \frac{1}{k_B T}  \sum_n| \langle n| {\cal O} | n\rangle|^2
\exp{(\frac{-E_n}{k_BT})}\\ \nonumber
&+\frac{2}{Z}   \sum_{n\neq m} \frac{|\langle n| {\cal O} | m\rangle|^2}{E_n-E_m} \exp{(\frac{-E_n}{
k_BT})}.\nonumber
\end{eqnarray}
Here $|n\rangle$ and $E_n$ with $n=1,\dots13$  are the eigenstates and eigenvalues of the Hamiltonians
${\cal H}^{l_i}$ obtained from eq.~\ref{H1} by the above described rotations to the laboratory system
and $F$ and $Z$ are the free energy and  partition sum.
Also in analogy to the magnetic susceptibility 
the first two contributions to $\chi(T)$
are referred to as Curie terms,
the last term as Van Vleck term. 
In general the Curie terms  depend strongly on the temperature 
and the Van Vleck term has a relatively weak temperature dependence. 

Using the magnetoelastic Hamiltonian eq. \ref{H3} the change in the elastic constants and the 
corresponding
strain
susceptibilities $\chi$  are given by:
\begin{eqnarray}
\label{H_sus1}
        \Delta c' = N g^2(\Gamma_3)  \frac{\partial <{\cal O}_{22}> }{\partial e_{2}} =
         N g^2(\Gamma_3) \chi(\Gamma_3)
\end{eqnarray}

and
\begin{eqnarray}
\label{H_sus2}
        \Delta c_{44} = N g^2(\Gamma_5)  \frac{\partial <{\cal O}_{xy}> }{\partial e_{xy}} =
         N g^2(\Gamma_5) \chi(\Gamma_5) .
\end{eqnarray}

\begin{figure} 
\begin{center}
\includegraphics[width=12cm,angle=0]{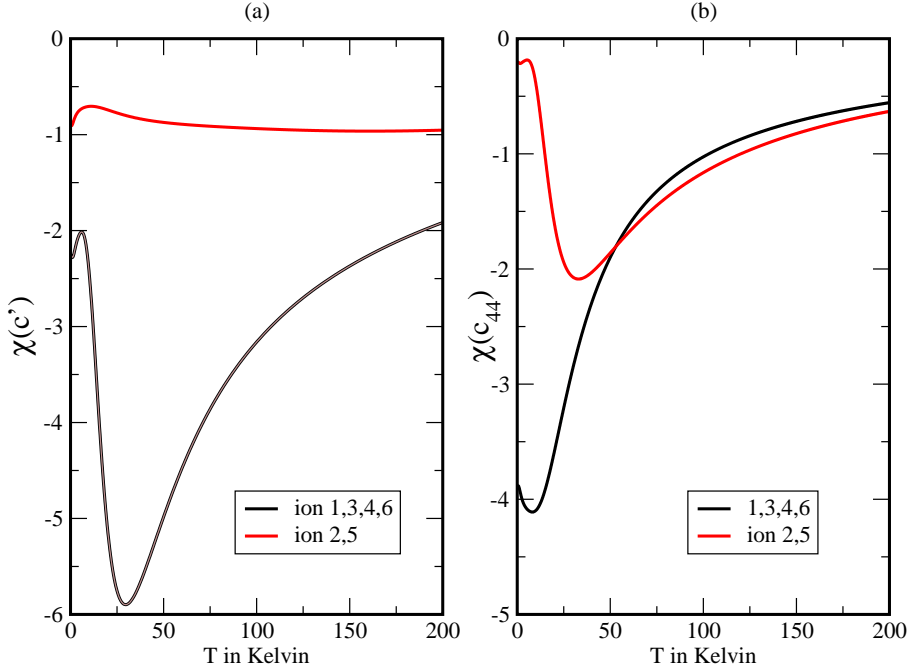}
\end{center}
\caption{(Color on line) Contributions to $\chi(c')$ (left panel) and contributions to
  $\chi(c_{44})$ (right panel) from the ions $l_i$ for $i=1,\dots 6$.}
\label{fig3}
\end{figure}

The strain susceptibilities are calculated for the six ions separately. The results are shown in Fig.~\ref{fig3}.
As can be seen from Fig.~\ref{fig3}a there are two distinct curves contributing to $\Delta c'$, one showing  a broad
minimum at about 30 K and distinct anomalies below 10 K and a
second one which varies little with temperature above 25 K.
We discuss the anomalies below 10 K in more detail in section \ref{anom}.
As shown in Fig.~\ref{fig3}b there are also two type of curves contributing to $\Delta c_{44}$ with minima at
10 K and 28 K. It may be worth noting that calculating $c_{55}$ or $c_{66}$ explicitly
instead  of $c_{44}$  gives the same results as shown
in Fig.~\ref{fig3} but with the role of the ions interchanged.
The same interchanging of ions is encountered when considering $(c_{33} -c_{13})/2$ 
instead of $c' $ .
This naturally reflects the cubic symmetry of the unit cell.

\begin{figure} 
\begin{center}
\includegraphics[width=12cm,angle=0]{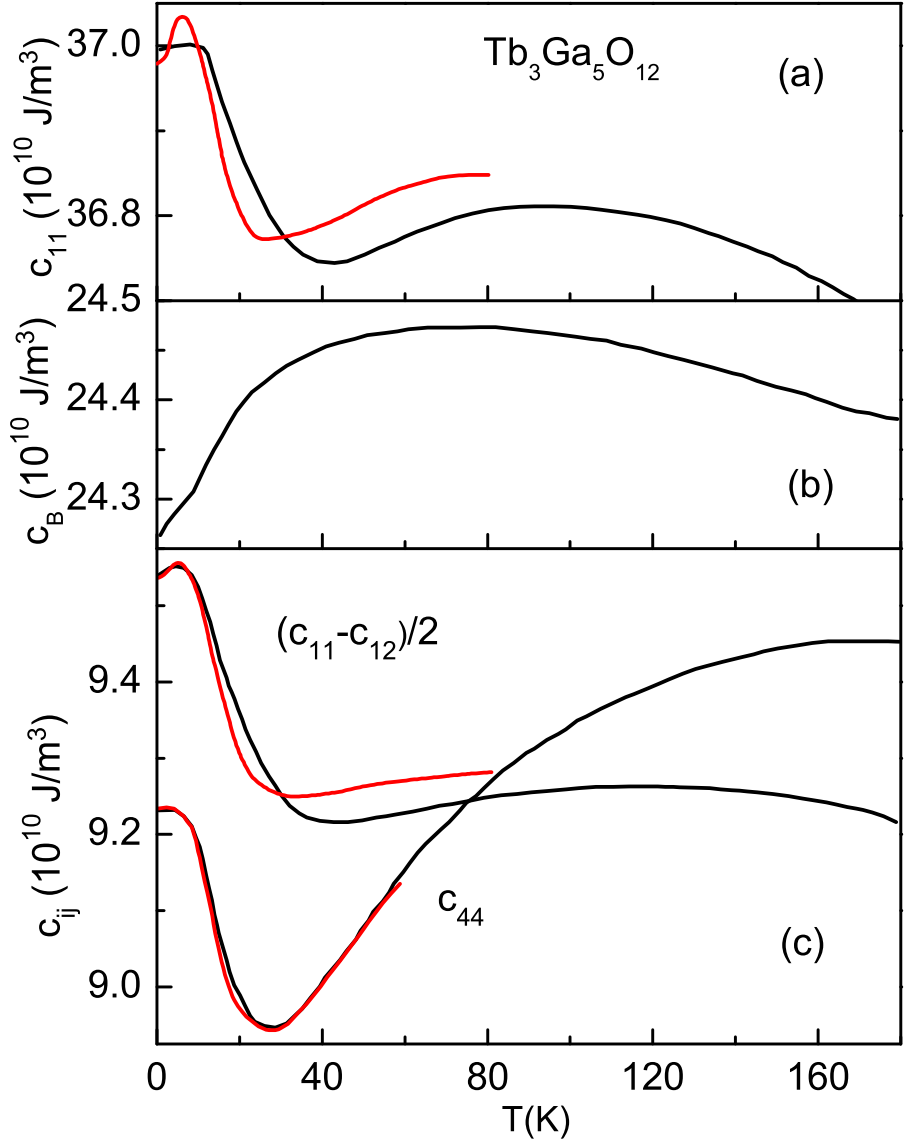}
\end{center}
\caption{(Color on line) Temperature dependence of the elastic constants
  $c_{11}$, $c_{B}$, $c'=(c_{11}-c_{12})/2$, and $c_{44}$. Experiment (black) and calculation (red). For
  the fit parameters and the background see Table I.}
\label{fig4}
\end{figure}

We fitted the curves  to the experimental data 
shown in Fig.~\ref{fig4}  averaging
over the contributions of the six different ions of Fig.~\ref{fig3}.
Since the two ions number 2, 5 and the four ions 1,3,4,6 give 
identical results for the elastic modes $c'$ and $c_{44}$
for brevity we henceforward denote the strain susceptibilities and the magneto-elastic
coupling constants 
with $\chi_2$ and $\chi_4$ and $g_2$ and $g_4$ respectively, using the multiplicity 2 and 4
as index.  

Therefore we get

\begin{eqnarray}
\label{H_araki1}
c' ={c'}_0 + \frac{N}{3}\left[ 2 (g_4(\Gamma_3))^2 {\chi}_{4}(\Gamma_3) + (g_2(\Gamma_3))^2 {\chi}_2(\Gamma_3) \right] 
\end{eqnarray}

\begin{eqnarray}
\label{H_araki2}
c_{44} =(c_{44})_0 + \frac{N}{3} \left[ 2 (g_4(\Gamma_5))^2 {\chi}_{4}(\Gamma_5) + (g_2(\Gamma_5))^2 {\chi}_2(\Gamma_5) \right]. 
\end{eqnarray}

For the background  we usually take the Varshny formula \cite{Varshny}.
For the low temperature region in which we are mostly interested, $c_0$ is almost constant.

The temperature dependence of the elastic constants $c'$, $c_{44}$, $c_{11}$, and $c_B$ are
given in Fig.~\ref{fig4}. The first three propagating modes were measured directly and
the non-propagating bulk modulus $c_B$ is calculated using the formula 

\begin{eqnarray}
\label{H_araki3}
c_B= (c_{11} + 2 c_{12})/3 =c_{11} -\frac{4}{3} c' 
\end{eqnarray}

All three propagating modes can be described quantitatively 
using eqs.~\ref{H_araki1},\ref{H_araki2}  and the corresponding strain susceptibilities 
of Fig.~\ref{fig3}. For the fit of the longitudinal $c_{11}$ mode we took the bulk 
modulus from eq.~\ref{H_araki3} and the calculated strain susceptibility 
for $c'$ with the same magneto-elastic coupling constants 
$g_{2,4}(\Gamma_3)$.   Especially the pronounced minimum at 30 K is given 
exactly with the calculated strain susceptibility for $c_{44}$. This mode is fitted 
particularly well.

\begin{figure} 
\begin{center}
\includegraphics[width=14cm,angle=0]{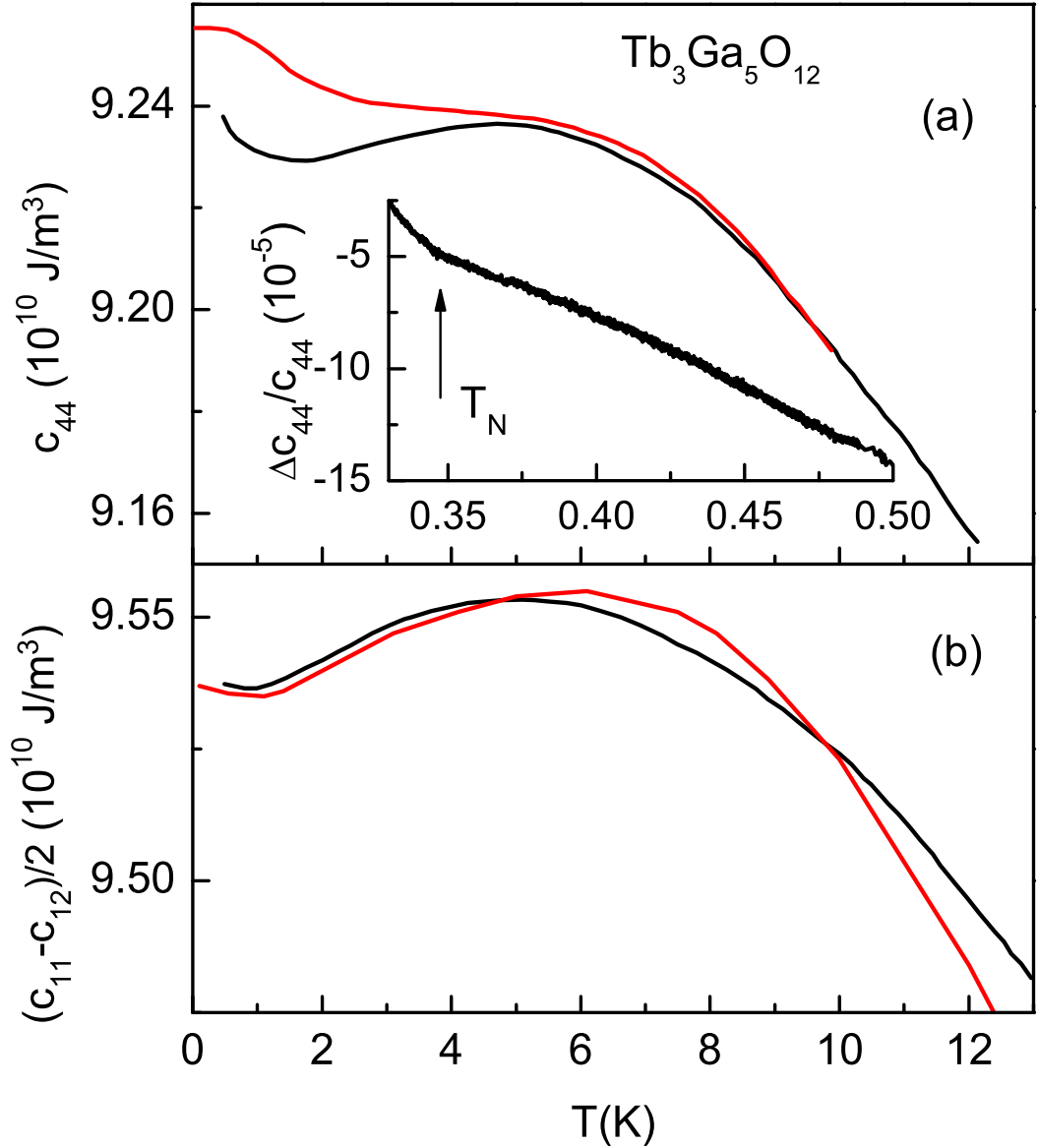}
\end{center}
\caption{(Color on line) Low temperature behaviour of the elastic constants. Experiment (black) and calculation (red). Inset shows a relative change of the elastic constant,  $c_{44}$, below 0.5 K, in the vicinity of the AFM ordering.}
\label{fig5}
\end{figure}

The fits for $c'$ and $c_{11}$ are less satisfactory. The main reason is
that the calculated minimum of $\chi_4(\Gamma_3)$ is at
30 K like $\chi_2(\Gamma_5)$ but experimentally
the minima of $c'$ and $c_{11}$ are at 40 K.
In addition $\chi_2(\Gamma_3)$ has little structure and is much smaller
than $\chi_4(\Gamma_3)$. Avoiding unrealistic large $g_2(\Gamma_3)$ for the high 
temperature fit, for low temperatures a good fit results in a 
neglect of $g_2$. The fits for $c'$ and $c_{11}$ are therefore very good
for  $T < 10$ K (see Fig.~\ref{fig5}) but  give only the salient features for higher temperatures 
(Fig.~\ref{fig4}).

We suspect that higher order magneto-elastic couplings (hexadecapole moment-strain coupling) had to
be considered for these modes. Such higher order couplings were introduced for a number of rare-earth 
compounds like PrSb, PrPb$_3$, PrNi$_5$ \cite{BL}.
The magneto-elastic coupling constants $g_2$ and $g_4$ 
used for the fit of the $c'$, $c_{11}$ and $c_{44}$ modes 
are given in Table \ref{table_fit}.

The temperature dependence of the bulk modulus $c_B$ shown also 
in Fig.~\ref{fig4} is  anomalous. Below 100 K it decreases continuously. 
This means that $c_B$ is also affected by the crystal field. 
Since $e_V$ and $c_B$ have $\Gamma_1$-symmetry this mode couples 
directly to the CEF Hamiltonian. Therefore we can write

\begin{eqnarray}
\label{H_Bruno2}
H_{me}(c_B)= G_4 e_V ({\cal O}_{40}+5{\cal O}_{44})  + G_6 e_V ({\cal O}_{60}-  21 {\cal O}_{66}) 
\end{eqnarray}

$G_4$ $G_6$ are coupling constants which can be determined by 
fitting the strain susceptibility deduced from eq.~\ref{H_Bruno2}, to the $c_B(T)$ curve
of Fig.~\ref{fig4}. This will be done together 
with the higher order susceptibility fits for 
for $c'$ and $c_{11}$ in later work.

\begin{table}
\begin{center}
\begin{tabular}{|c|c|c|c|c|c|c|c|}
\hline
$ $ & 
$g_2(\Gamma_5)$ & 
$g_4(\Gamma_5)$ & 
$g_2(\Gamma_3)$ & 
$g_4(\Gamma_3)$ & 
$g_F$ & 
$g_{CM}$ &
$c_0(T=0)$ 
\\
\hline
  $c_{44}(\Gamma_5)$ & 
  193 K   &  98 K
   & & &
 81 K &  55 K
  & 
   $9.74 \cdot 10^{10}$ \\
\hline
  $c'(\Gamma_3)=(c_{11}-c_{12})/2$ & 
   & 
   & 0 K
         & 
    114 K & 
 & &      $9.71 \cdot 10^{10}$ \\ 
\hline
  $c_{11}$ & 
   & 
   & 0 K 
  & 
 114 K & 
    & & \\
\hline
\end{tabular}
\end{center}
\caption{Magneto-elastic coupling constants, $g_i$, and background elastic constants, $c_0$,
  in $J/m^3$}
\label{table_fit}
\end{table}

\section{ Elasticity due to the ground-state quasi-doublet and the resonant spin-phonon effects} 
\label{anom}

As seen in the temperature dependence of the various elastic constants in Fig.~\ref{fig4} there are, 
apart from the strong anomalies around 30 K, also weaker extrema for $T < 4$ K. As demonstrated above
these structures are well reproduced by the calculation (see Fig.~\ref{fig5}). Inset of Fig.~\ref{fig5} shows a change of the slope in the $c_{44}$ at the AFM ordering \cite{Kama}. Note, that it is a rather unusual feature. One might more likely expect an anomaly at $T_{N}$ in the acoustic properties of a longitudinal mode. In any case the small effect at $T_{N}$ on $c_{44}$ shows that the antiferromagnetic ordering has a negligible effect on $c_{44}(T,B)$ displayed in the Figs.~\ref{fig8}, \ref{fig9}, \ref{fig10}.

\begin{figure} 
\begin{center}
\includegraphics[width=14cm,angle=0]{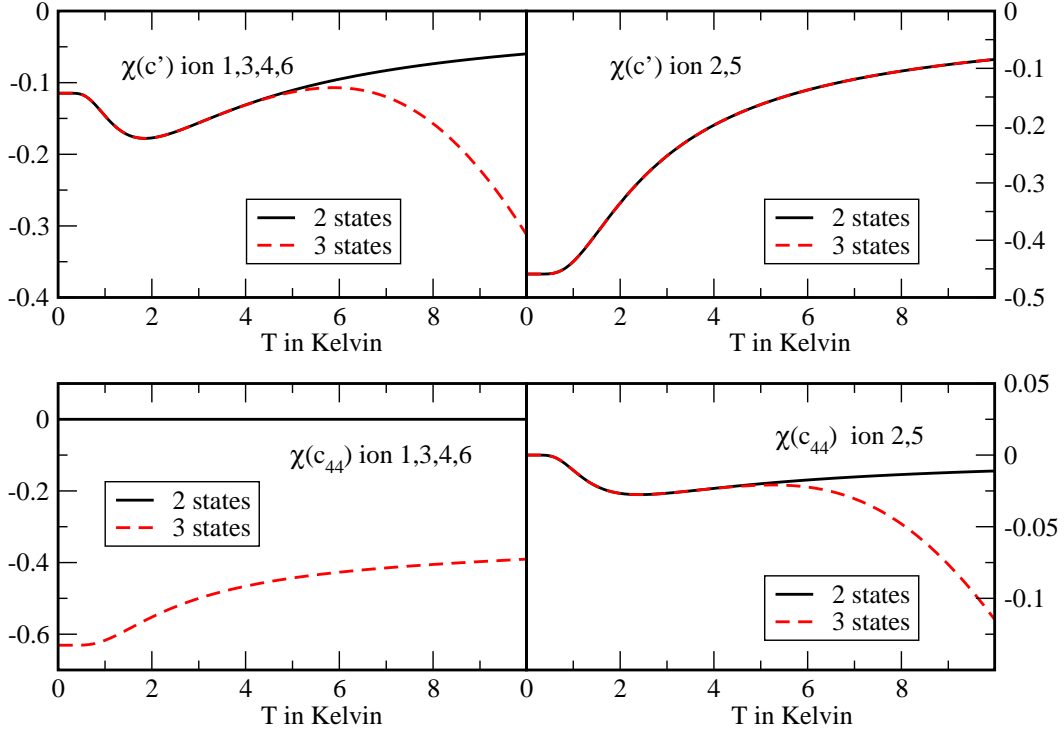}
\end{center}
\caption{(Color on line) Contributions of the lowest lying states to  $c' = (c_{11}-c_{12})/2$ and $c_{44}$. }
\label{fig6}
\end{figure}

In the following we demonstrate  that the low-temperature anomalies are mainly due to the quadrupolar 
couplings within the lowest states alone.
For this the strain susceptibilities were calculated using 
a reduced ensemble of the lowest two and the lowest three states.
The results are shown in Fig.~\ref{fig6}.  Comparing with Fig.~\ref{fig3} one sees that 
the  low-temperature anomalies, similar to the ones shown in Figs.~\ref{fig4},\ref{fig5}  can be found also in the 
calculation within the reduced ensemble. The strong minimum at 30 K is present  only with
three and higher states included. 

One should note, however, that the overall height of the strain susceptibility cannot be accounted for by using the lowest states alone,
since matrix elements between the intermediate states substantially contribute 
to the strain susceptibilities also close to zero temperature. This is mainly due to the van Vleck
contribution to the susceptibility. The partition function at low temperatures is, of course, well described 
by the lowest states alone.

The zero field splitting ($B$ = 0) of the quasi doublet (0,1) is 3.7 K 
as discussed in section \ref{CEF} and Ref.~\cite{us3}. This splitting is too large 
for a study of resonant phonon effects with coherent sound waves. 
The splitting corresponds to microwave phonons of 77 GHz. However in a 
thermal conductivity experiment this splitting was observed as a 
minimum at 0.52 K (see Ref.~\cite{Inyushkin2010}). It was interpreted as a resonant phonon
scattering  process for a two-level system.
Magnetic fields close the gap of the quasi doublet only with the additional 
energy level 3 for $B$ = 9 T ($[110]_c$-direction) and at $B$ = 19 T
($[100]_c$-direction) 
as shown in  Fig.~\ref{fig2}
and observed in ESR and magnetization experiments \cite{us3,Guillot}. 
 Therefore sound attenuation experiments in magnetic fields 
in TGG do not provide ideal conditions for studying resonant spin-phonon 
interaction for a two-level system. 
Likewise the theory of the phonon-Hall effect in TGG should not treat only the quasi 
doublet but should take at least the lowest three states into account. In thermal conductivity the resonant interaction leads to strong thermal resistance \cite{Inyushkin2010}. 

\begin{figure} 
\begin{center}
\includegraphics[width=12cm,angle=0]{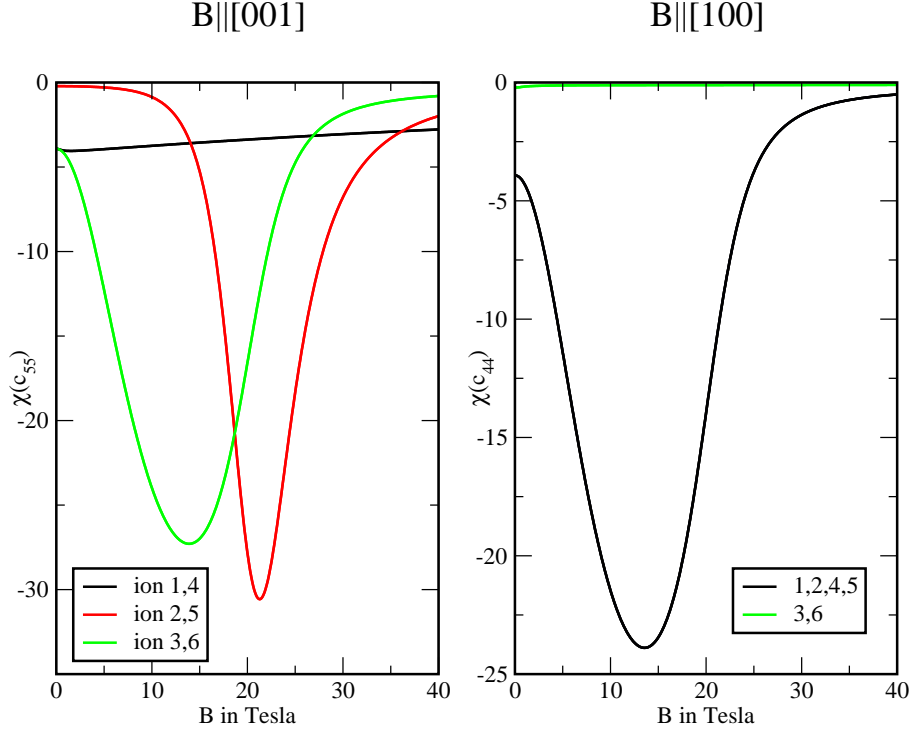}
\end{center}
\caption{(Color on line) Contributions to $\chi(c_{55})$  for field in $[001]_c$ direction (Faraday geometry) and contributions to $\chi(c_{44})$ for field in $[100]_c$ direction (CM geometry) from the ions $l_i$ for $i=1,\dots 6$.}
\label{fig7}
\end{figure}

\section{Elastic constants in magnetic field}

We investigate elastic constants in magnetic fields for different geometries. 
One is the so-called Faraday geometry, another one the Cotton-Mouton-Voigt
geometry. 
We use the same formalism  to calculate the strain susceptibilities in
the presence of a magnetic field. Now there are two independent 
directions  given by the direction of the sound wave and 
of the magnetic field. 

\subsection{Faraday geometry}

\begin{figure} 
\begin{center}
\includegraphics[width=12cm,angle=0]{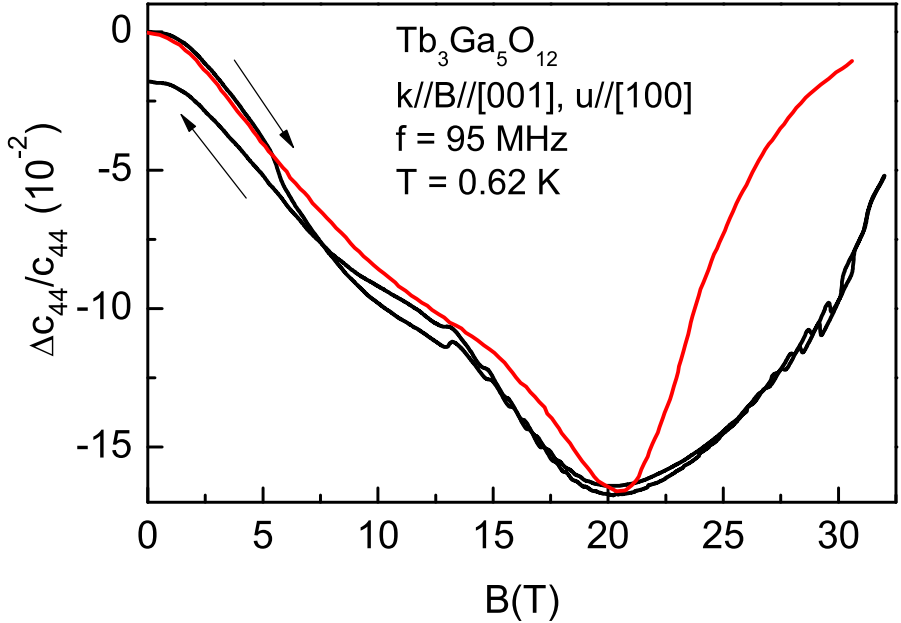}
\end{center}
\caption{(Color on line) Faraday configuration: Magnetic field dependence of the elastic constant,
  $c_{44}$, experiment (black, field sweeps up and down are shown) and calculation (red).}
\label{fig8}
\end{figure}

This geometry was already discussed qualitatively in Ref.~\cite{us2}. 
In Fig.~\ref{fig7} we give the calculated strain susceptibility for 
$c_{44}$ with $k||B||[001], u||[100]$ involving all 6 ions. The strain
  susceptibilities $\chi(c_{55})$ in this case are degenerate for the three pairs 
(1,4), (2,5) and (3,6). If we assume the same magneto-elastic coupling
  constant for the three pairs we get the averaged susceptibility $\chi_{av}$, 
 shown in Fig.~\ref{fig8}. This $\chi_{av}$ has the same form as the
experimentally observed one also shown in Fig.~\ref{fig8}. 
Therefore we take this $\chi_{av}$ to fit the experiment using the formula

\begin{eqnarray}
\label{Bruno3}
\Delta c_{44} = - g_{F}^2 N \chi_{av}
\end{eqnarray}

This gives a magneto-elastic coupling constant $g_F = 80.6$ K. Note that eq.~\ref{Bruno3}
gives the minimum exactly at the same field of $20.3$ T as the experiment. 
The deviation at higher fields may be due to the averaging over the three pairs of
ions. It could also be due to the magneto-caloric effect, as calculated in Ref.~\cite{us2}. 

\begin{figure} 
\begin{center}
\includegraphics[width=12cm,angle=0]{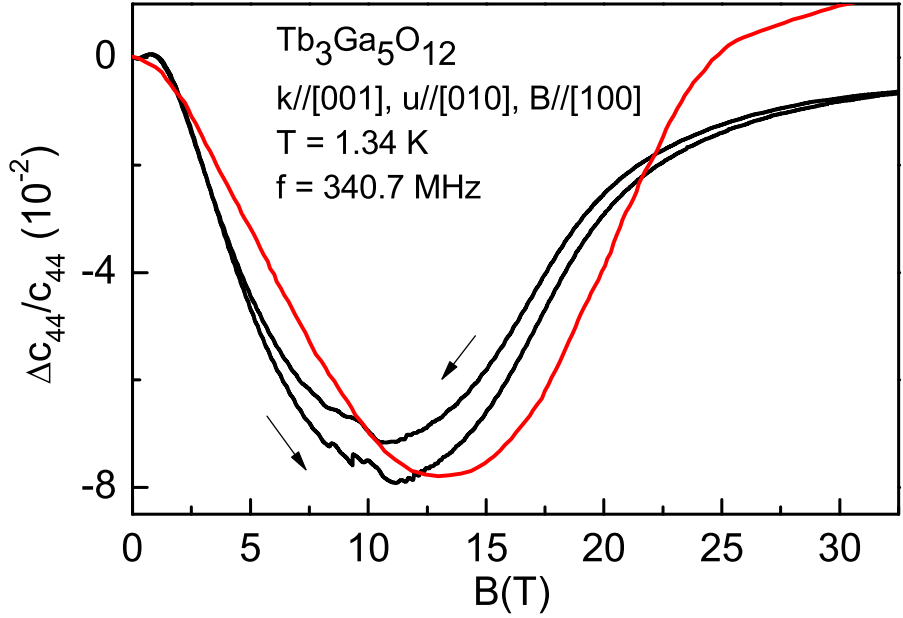}
\end{center}
\caption{(Color on line) Voigt configuration: Magnetic field dependence of the elastic constant,
  $c_{44}$, experiment (black, field sweeps up and down are shown) and calculation (red).}
\label{fig9}
\end{figure}

\subsection{Cotton-Mouton geometry}

Here we measured the elastic constant $c_{44}$ for $k||[001]$, $u||[010]$ and
$B||[100]$. 
The experimental result is shown in Fig.~\ref{fig9} for $T = 1.34$ K and frequency of 340.7 MHz. The hysteresis for field increase and decrease is
 probably due to some heating effect, so the increasing one is closer to the
 given temperature. The calculated strain susceptibility $\chi(c_{44})$ is also shown in
 Fig.~\ref{fig9}. In this geometry four ions give identical results (1,2,4,5) whereas 
 the remaining two (3,6) provide a negligibly small contribution.  Therefore we obtain one
 coupling constant  $g_{CM}$ with an additional factor $4/6$ for the strain susceptibility
(see Fig.\ref{fig7}).

In the case of the Cotton-Mouton geometry  the minima of measured and
calculated curves differ slightly. The measured minimum is at 11 T and the
calculated one at 13.6 T. Since the form of the curves are very similar they 
are just shifted by 2 T from each other. The magneto-elastic coupling constant
from the fit gives $g_{CM} = 55.2$ K. 

The important point is that in the Faraday geometry the minimum of the
$c_{44}$-mode versus field is at 20.3 T in excellent agreement with our
calculation, for the CM-geometry it is at a much lower field of 10 - 13 T 
in agreement with our calculation too. This gives strong support for the CEF scheme
proposed in Ref~\cite{Guillot}  and also used for magnetic and ESR investigation in
Ref.~\cite{us3}.  The minimum for the Faraday geometry is due to the crossover of the 
lowest two energy levels at 20 T as seen in Fig.~\ref{fig2}. The broad minimum in the CM
geometry is at a lower field because these ions experience a smaller field 
as seen from the calculated strain susceptibilities of Fig.~\ref{fig9}.

Unfortunately we do not have results for the other Cotton-Mouton geometry:
$k||[001]$, $ u|| B|| [100]$. This would have allowed us to investigate the influence
of asymmetric strain contributions, the so-called rotationally invariant
magneto-elastic contribution \cite{BL}.

\subsection{Coupling constants discussion}

In Table 1 we list the various magneto-elastic coupling constants from the
temperature dependence and from the magnetic field dependence of the elastic
constants. For the $c_{44}$ mode which was investigated as a function of
temperature and magnetic field we find the coupling constants all in the range
from 50 to 200 K. Of course, the various $g(\Gamma_5)$ have not to be
exactly the same, since the local coordinate systems 
differ with respect to the magnetic field direction. 
For the $c'$ mode the negligible coupling constant, $g_2$, and the sizable 
value for $g_4$ describe the low temperature properties quite well 
as seen in Fig.~\ref{fig5}. The slight disagreements for higher temperatures are 
due to the different minima positions observed experimentally (40 K) and calculated (30 K). 
Possible further reasons for the disagreement were given in chapter \ref{MEC}.

In this paper we investigated magneto-elastic couplings with single ion
effects. We neglected two ion effects like e.g. a direct quadrupole-quadrupole
interaction (see ref.\cite{BL} section 5.3) for the following reasons: 
For c'(T) such 2-ion effects do not improve the fit and for c$_{44}$(T) the fit is
excellent  without this additional coupling. For c$_{44}$(B) the inclusion of two
ion 
effects is rather difficult because of the strong field dependence of the 
strain susceptibility of the 6 different ions (Fig.\ref{fig7}). 
One had to introduce at least 2 - 3 more coupling constants which makes a fit meaningless. 

\section{ACOUSTIC COTTON-MOUTON EFFECT IN TGG } 

In previous papers \cite{us1,us2} we have studied the acoustic Faraday
effect in TGG.  Here we investigate the Cotton-Mouton-Voigt effect in this material. 
In the Faraday effect $B || \vec k$ which leads to a rotation of the polarization direction.

In the Cotton-Mouton effect $B\perp \vec k$ which leads to birefringence. 
The velocities for $B || \vec u$ and $B \perp \vec u$ are different
so we encounter a phase change $\frac{\Phi(B)}{L}$ for  $k||[001], B||[100], u||[110]$ (see Ref.~\cite{BL})
given by 

\begin{eqnarray}
\label{Bruno4}
\frac{\Phi(B)}{L}= \omega \left( \frac{1}{v_{100}} -\frac{1}{v_{010}}\right).
\end{eqnarray}

\begin{figure} 
\begin{center}
\includegraphics[width=10cm,angle=0]{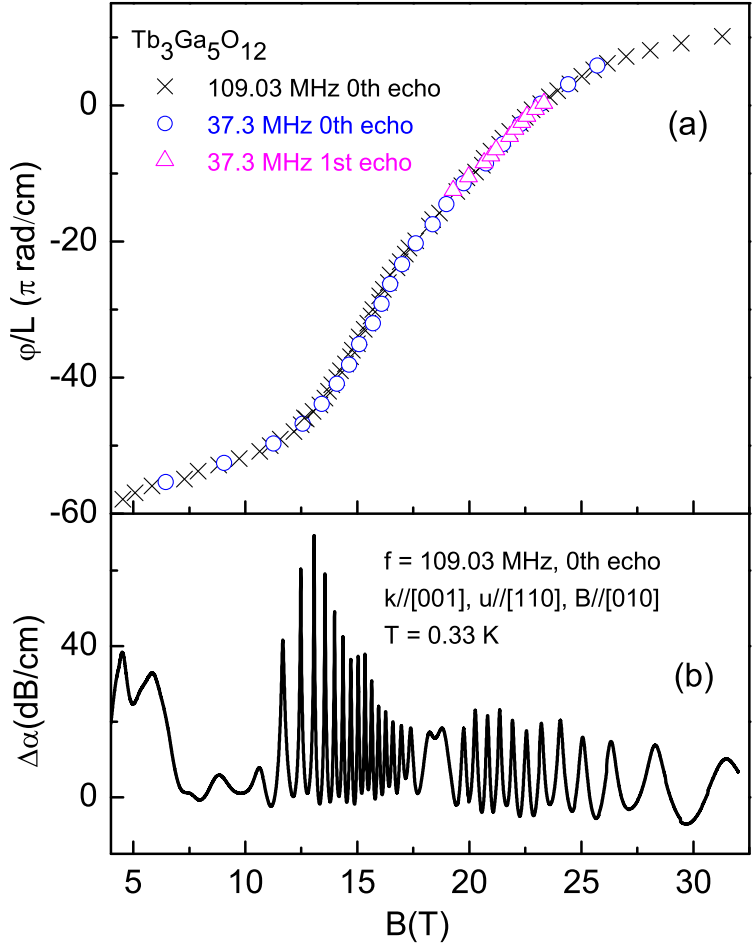}
\end{center}
\caption{(Color on line) (a) Phase $\Phi/L$ versus magnetic field for various ultrasound signals. The frequencies and echoes are normalised to this curve using eq. \ref{Bruno4}.
(b) Echo amplitude as a function of $B$, $f$ = 109 MHz, $T$ = 0.33 K.}
\label{fig10}
\end{figure}

A typical example of the amplitude modulation of a given ultrasonic echo as a 
function of magnetic field is exhibited in Fig.~\ref{fig10}b.

A linearly polarized wave with $u || [110]$ changes in the field to elliptically 
polarized and after a phase change of $\pi/2$ to circularly polarized, followed by 
elliptical polarization and finally linearly polarization orthogonal to the original 
linearly polarization $u || [1\bar10]$.  The phase difference between subsequent maxima and minima 
is therefore $\pi$. 

Different echoes with different $L$  and with different frequencies give a unique
$\Phi(B)/L$ plot, normalised to one frequency $\omega  = 2\pi f $ and one length $L$, 
also shown in Fig.~\ref{fig10}a.

We notice that the linear frequency dependence of eq.~\ref{Bruno4} and the dependence 
on the travel distance $L=L_0(2n+1)$ ($n$ is the echo number) is strictly observed.  With eq.~\ref{Bruno4} one 
could in principle calculate the  $\Phi(B)/L$ by using the measured 
velocity curves $v_{100}(B)$ and $v_{010}(B)$. As pointed out above the mode $v_{100}$ was not measured, but the Faraday mode should give the same $B-$dependence for symmetric strains. The $v(B)$ curves should be measured at the same temperature, however. In addition both $v(B)$ curves exhibit similar forms and have minima at 11 T and 20 T respectively. This leads to extremely sensitive $\phi(B)$ dependence. Therefore only qualitative fits are  possible. They give the right order of magnitude however. 

This $\Phi(B)$- behaviour has to be compared with another experiment performed in  CeAl$_2$ ~\cite{BLCL,BL}. Here the two velocity modes have opposite field
dependencies  and the agreement of the measured $\Phi/L$ curve with eq.~\ref{Bruno4} is perfect.

\section{Conclusion}

Tb-Ga-Garnet with its many unusual properties has been investigated with ultrasound as 
a function of temperature, down to 0.3 K and at high magnetic fields. The local $D_2$ symmetry 
of the six inequivalent Tb$^{3+}$ ions leads to pronounced crystal-field effects in 
magnetization \cite{Guillot}, ESR \cite{us3}, and elastic constants investigated in this work. 
For the magneto-elastic interaction and the resulting phonon effects it was important 
to transform the CEF Hamiltonian to the laboratory system, where the elastic 
constants and the magneto-elastic Hamiltonian can be described in the usual cubic symmetry.

The temperature dependence of the elastic constants can be described quantitatively, where 
the important $c_{44}$ mode is especially well fitted. The magnetic field dependence of the 
c$_{44}$ mode provided crucial tests for the CEF –scheme. We found very good agreement for the 
Faraday geometry with a minimum at 21 T and a small discrepancy for the Cotton-Mouton geometry with 
a minimum experimentally at 11 T and by CEF calculation at 13 T. Besides the acoustical Faraday 
effect \cite{us2} we showed analogous amplitude modulations for the acoustical Cotton-Mouton effect. 
The frequency dependence (linear in $\omega$)is observed.

This detailed theoretical and experimental investigation gives a foundation for a realistic treatment of 
other effects, such as the phonon Hall effect \cite{Strohm2005,Inyushkin2007}  and higher order magneto-elastic effects.
\newpage
\section{Acknowledgment}
We acknowledge the support of the HFML-RU/FOM and HLD at HZDR, members of the European Magnetic Field Laboratory (EMFL). This work was partly supported by the EuroMagNET II Project financed by the European Community under Contract 228043. We thank support by the Strategic Young Researcher Overseas Visits Program for Accelerating Brain Circulation from the Japan Society for the Promotion of Science (JSPS). We thank Peter Wyder for helpful discussions.

\section{Appendix}
\subsection{Summary of crystal field parameters, Coordinate systems and Tensor Operators}

In the following table we list the crystal field parameters $a_{ij}$ of TGG in cm$^{-1}$  as found in 
\cite{Guillot}. 

\begin{eqnarray}
\begin{array}{|c|c|c|c|c|c|c|c|c|}
\hline
a_{20} & 
a_{22} & 
a_{40} & 
a_{42} & 
a_{44} & 
a_{60} & 
a_{62} & 
a_{64} & 
a_{66} \\
\hline
  -81.0 & 
  169.0 & 
-2163.0 & 
  249.0 & 
  945.0 & 
  677.0 & 
 -155.0 & 
 1045.0 & 
   -4.0 \\
\hline
\end{array}
\label{A1}
\end{eqnarray}\\

The connections to the $b_{ij}$ used in eq.~\ref{H1} is given by 
\begin{eqnarray}
&b_{2j}=  \frac{a_{2j} \alpha_J}{f_{2j}} & \ \ j=0,2\\\nonumber
&b_{4j}=  \frac{a_{4j} \beta_J}{f_{4j}} & \ \ j=0,2,4\\\nonumber
&b_{6j}=  \frac{a_{6j} \gamma_J}{f_{6j}} & \ \ j=0,2,4,6
\label{A2}
\end{eqnarray}

with 
\begin{eqnarray}
\label{A3}
&\alpha_J=\frac{-1}{99}\\
&\beta_J=\frac{2}{16335}\\
&\gamma_J=\frac{-1}{891891}
\end{eqnarray}


and the $f_{ij}$ given in  eq.\ref{A4}. 

\begin{eqnarray}
\begin{array}{|c|c||c|c|} 
 \hline 
 f_{20} & 2               & f_{21} &1/\sqrt{6}\\
 f_{22} & 2/\sqrt{6}      & f_{41} &2/\sqrt{5}\\
 f_{40} & 8               & f_{43} &2/\sqrt{35}\\
 f_{42} & 8/\sqrt{40}     & f_{61} &\sqrt{32/21}\\
 f_{44} & 8/\sqrt{70}     & f_{63} &8/\sqrt{105}\\
 f_{60} & 16              & f_{65} &8/\sqrt{693}\\
 f_{62} & 16/\sqrt{105}  & & \\
 f_{64} & 16/\sqrt{126}  & & \\
 f_{66} & 16/\sqrt{231}  & & \\
 \hline 
 \end{array}
\label{A4}
\end{eqnarray}

The notation for the tensor operators in the literature is far from being
unique. We follow the notation of 
M.T. Hutchings \cite{Hutchings64} and P.A. Lindg\r ard, O. Danielsen  \cite{Lind},
but for reasons of clarity we also list
the explicit form of the tensor operators ${\cal O}_{l\pm m} $used in our work
in eq. \ref{A5}, \ref{A6}, \ref{A7}.  Note that the 
${\cal O}_{l\pm m}= \tilde{\cal O}_{l\pm m} f_{lm}$,  where $\tilde{\cal
  O}_{l\pm m}$ with $m=0,1 \dots l$ are the 
Racah operator equivalents given in Table 1 of \cite{Lind}.
In the following $\{A,B\}$ denotes the anticommutator, $J_i$ with $i=x,y,z$ the compnents of the angular momentum, 
$J$ the total angular momentum and $J_\pm=J_{x}\pm i J_{y}$.

\begin{eqnarray}
\begin{array}{|c|c|} 
\hline
{\cal O}_{2,0}   &   3 J_{z}^2 - J(J+1)  \\ 
{\cal O}_{2,+1}   &   -\frac{1}{4} \{ J_{+} ,J_{z} \} \\ 
{\cal O}_{2,-1}   &   \frac{1}{4} \{ J_{-} ,J_{z} \} \\ 
{\cal O}_{2,+2}   &  \frac{1}{2} J_{+}^2  \\ 
{\cal O}_{2,-2}   &  \frac{1}{2} J_{-}^2  \\ 
\hline
\end{array}
\label{A5}
\end{eqnarray}
\begin{eqnarray}
\begin{array}{|c|c|} 
\hline
{\cal O}_{4,0}   &   35 J_{z}^4 -30 J_{z}^2 J(J+1) +25 J_{z}^2 -6 J(J+1) +3 J^2(J+1)^2  \\ 
{\cal O}_{4,+1}  &  -\frac{1}{4} \{ 7 J_{z}^3-(3 J(J+1) + 1) J_z ,J_{+} \} \\ 
{\cal O}_{4,-1}  &  \frac{1}{4} \{ 7 J_{z}^3-(3 J(J+1) + 1) J_z ,J_{-} \} \\ 
{\cal O}_{4,+2}  &  \frac{1}{4} \{ 7 J_{z}^2-J(J+1)-5,J_{+}^2 \} \\ 
{\cal O}_{4,-2}  &  \frac{1}{4} \{ 7 J_{z}^2-J(J+1)-5,J_{-}^2 \} \\ 
{\cal O}_{4,+3}  &  -\frac{1}{4} \{ J_{+}^3,J_z \} \\ 
{\cal O}_{4,-3}  &  \frac{1}{4} \{ J_{-}^3,J_z \} \\ 
{\cal O}_{4,+4}   &  \frac{1}{2} J_{+}^4  \\ 
{\cal O}_{4,-4}   &  \frac{1}{2} J_{-}^4  \\ 
\hline
\end{array}
\label{A6}
\end{eqnarray}
\begin{eqnarray}
\begin{array}{|c| c|} 
\hline  
{\cal O}_{6,0}  &   231 J_z^6 - 315 J_z^4 J(J+1) +735 J_z^4 +105 J_z^2
J^2(J+1)^2 -525 J_z^2 J(J+1)\\ 
& +294 J_z^2 -5 J^3(J+1)^3 + 40 J^2(J+1)^2 -60 J(J+1)\\
{\cal O}_{6,+1}  &    -\frac{1}{4} \{ 33 J_z^5 -(30 J(J+1) -15) J_z^3 +
(5 J^2(J+1)^2 - 10 J(J+1) +12)J_z , J_{+}\} \\
{\cal O}_{6,-1}  &    \frac{1}{4} \{ 33 J_z^5 -(30 J(J+1) -15) J_z^3 +
(5 J^2(J+1)^2 - 10 J(J+1) +12) J_z , J_{-}\} \\
{\cal O}_{6,+2}  &    \frac{1}{4} \{ 33 J_z^4 -(18 J(J+1) +123) J_z^2 +
J^2(J+1)^2 +10 J(J+1) +102 , J_{+}^2\} \\
{\cal O}_{6,-2}  &    \frac{1}{4} \{ 33 J_z^4 -(18 J(J+1) +123) J_z^2 +
J^2(J+1)^2 +10 J(J+1) +102 , J_{-}^2\} \\
{\cal O}_{6,+3}  &   -\frac{1}{4} \{ 11 J_z^3 -3J(J+1) J_z-59 J_z , J_{+}^3\} \\
{\cal O}_{6,-3}  &   \frac{1}{4} \{ 11 J_z^3 -3J(J+1) J_z-59 J_z , J_{-}^3\} \\ 
{\cal O}_{6,+4}  &   \frac{1}{4} \{ 11 J_z^2 -J(J+1) -38, J_{+}^4\} \\
{\cal O}_{6,-4}  &   \frac{1}{4} \{ 11 J_z^2 -J(J+1) -38, J_{-}^4\} \\
{\cal O}_{6,+5}  &  -\frac{1}{4} (J_{+}^5 J_z + J_z J_{+}^5) \\ 
{\cal O}_{6,-5}  & \frac{1}{4} (J_{-}^5 J_z + J_z J_{-}^5) \\ 
{\cal O}_{6,+6}  &  \frac{1}{2} J_{+}^6  \\ 
{\cal O}_{6,-6}  &  \frac{1}{2} J_{-}^6  \\ 
\hline
\end{array}
\label{A7}
\end{eqnarray}

For $l=6, m=6,4,2$ for $l=4, m=4,2$  and  for $l=2, m=2$ 
we also employ  the notation 
 ${\cal O}_{lm} = {\cal O}_{l+m}+ {\cal O}_{l-m} $
which is redundant of course, but since it is widely used, in particular 
in \cite {Guillot} we nonetheless adopt it in eq.~\ref{H1} and
whenever there is no need to introduce ${\cal O}_{l+m}$ and ${\cal O}_{l-m} $
separately.
 As it is common we also introduce the
 operators ${ \cal O}_{xy},{ \cal O}_{xz}$ and ${ \cal O}_{yz} $  defined as
 the anticommutator  ${ \cal O}_{ij} =\frac{1}{2} \{J_i, J_j \} $ 
 with $ i,j=x,y,z $  
 of the angular momentum operators  $J_x,J_y,J_z.$    

Next we list the transformations from  the
local systems $l_{i}, i=1,\dots 6$ 
to the laboratory system. 
The first column gives the transformation
matrices, column 2,3,4 give the unit vectors in the
local systems and the last three columns give the Euler angles in the notation
of \cite{Edmonds}. 

\begin{eqnarray}
\begin{array}{|l|l|l|l|l|l|l|} 
\hline 
Matrix & (e_x)_{local} & (e_y)_{local} &(e_z)_{local}&\alpha & \beta& \gamma\\ 
   \hline 
 R^1= 
\left(
\begin{array}{lll} 
    0                 &  \frac{1}{\sqrt 2 }                  & \frac{1}{\sqrt 2 }  \\
    0                 & - \frac{1}{\sqrt 2 }                 & \frac{1}{\sqrt 2 } \\
    1                 &   0                                  & 0
\end{array}
\right)
                    & [001]_c & [1\bar{1}0]_c  &[110]_c&  \alpha_1 = 0&
\beta_1  = \frac{\pi}{2}&
\gamma_1 = \frac{3\pi}{4}\\ 
   \hline 
 R^{2}= 
\left(
\begin{array}{lll} 
    0 & - \frac{1}{\sqrt 2 }      &  \frac{1}{\sqrt 2 }  \\
    1 &   0                       &  0                \\
    0 &  \frac{1}{\sqrt 2 }       &   \frac{1}{\sqrt 2 } .
\end{array}
\right)
                    & [010]_c & [\bar{1}01]_c  &[101]_c& \alpha_2 = \frac{\pi}{2}&
\beta_2  = \frac{\pi}{4}&
\gamma_2 = \pi\\
   \hline 
                  
 R^{3}= 
\left(
\begin{array}{lll} 
   1  &                     0  &   0 \\
   0  & \frac{1}{\sqrt 2}      &   \frac{1}{\sqrt 2}\\
   0  & -\frac{1}{\sqrt 2}     &   \frac{1}{\sqrt 2 } 
\end{array}
\right)
                    & [100]_c & [01\bar{1}]_c  &[011]_c& \alpha_3 = \frac{3\pi}{2}&
\beta_3  = \frac{\pi}{4}&
\gamma_3 = \frac{\pi}{2}\\
   \hline 
   \hline 
 R^4= 
\left(
\begin{array}{lll} 
    0 &   \frac{1}{\sqrt 2 } & -\frac{1}{\sqrt 2 } \\
    0 &  \frac{1}{\sqrt 2 } & \frac{1}{\sqrt 2 }\\
    1 &    0                & 0
\end{array}
\right)
                    & [001]_c & [110]_c  &[\bar{1}10]_c& \alpha_4 = 0&
\beta_4  = \frac{\pi}{2}&
\gamma_4 = \frac{\pi}{4}\\
   \hline 
 R^{5}= 
\left(
\begin{array}{lll} 
    0  & \frac{1}{\sqrt 2 } & \frac{1}{\sqrt 2 } \\
    1  & 0                  &  0 \\
    0  & \frac{1}{\sqrt 2 } &  - \frac{1}{\sqrt 2 } .
\end{array}
\right)
                            & [010]_c & [101]_c  &[10\bar{1}]_c& \alpha_5 = \frac{\pi}{2}&
\beta_5  = \frac{3\pi}{4}&
\gamma_5 = \pi\\
   \hline 
                  
 R^{6}=
 \left(
\begin{array}{lll} 
     1                & 0                   & 0 \\
   0& \frac{1}{\sqrt 2} &    -\frac{1}{\sqrt 2 } \\
   0& \frac{1}{\sqrt 2} &    \frac{1}{\sqrt 2 } 
\end{array}
\right)

                     & [100]_c & [011]_c  &  [0 \bar{1}1]_c &\alpha_6 =\frac{\pi}{2}&
\beta_6  = \frac{\pi}{4}&
\gamma_6 = \frac{3 \pi}{2}\\
   \hline 
\end{array}
\label{table1}
\end{eqnarray}

Using the appropriate representations of the rotation operators $D(\alpha,\beta,\gamma)= \exp{(i\gamma J_Z)} \exp(i\beta J_y) \exp(i\alpha J_Z )$ 
for angular momenta j=2,4,6 as given e.g. in eq.(4.1.12) of ref.\cite{Edmonds}   the rotated Hamiltonians ${\cal H}^{l_i}$ take the form
\begin{eqnarray}
            {\cal H}^{l_i}=               & b^{l_i}_{20} {\cal O}_{20}
		                      +b^{l_i}_{2,-1} {\cal O}_{2,-1}
		                      +b^{l_i}_{2,+1} {\cal O}_{2,+1}
		                       +b^{l_i}_{2,+2} {\cal O}_{2,+2}
		                       +b^{l_i}_{2,-2} {\cal O}_{2,-2}
                                       +b^{l_i}_{40} {\cal O}_{40}
                                       +b^{l_i}_{4,-1} {\cal O}_{4,-1}\\\nonumber
                                      +&b^{l_i}_{4,+1} {\cal O}_{4,+1}
                                       +b^{l_i}_{4,-2} {\cal O}_{4,-2}
                                       +b^{l_i}_{4,+2} {\cal O}_{4,+2}
                                       +b^{l_i}_{4,-3} {\cal O}_{4,-3}
                                       +b^{l_i}_{4,+3} {\cal O}_{4,+3}
		                       +b^{l_i}_{44} {\cal O}_{44}
                                       +b^{l_i}_{60} {\cal O}_{60}\\\nonumber
                                       +&b^{l_i}_{6,-1} {\cal O}_{6,-1}
                                       +b^{l_i}_{6,+1} {\cal O}_{6,+1}
                                       +b^{l_i}_{6,-2} {\cal O}_{6,-2}
                                       +b^{l_i}_{6,+2} {\cal O}_{6,+2}
                                       +b^{l_i}_{6,-3} {\cal O}_{6,-3}
                                       +b^{l_i}_{6,+3} {\cal O}_{6,+3}
		                       +b^{l_i}_{64} {\cal O}_{64}  \\\nonumber
                                       +&b^{l_i}_{6,-5} {\cal O}_{6,-5}
                                       +b^{l_i}_{6,+5} {\cal O}_{6,+5}
		                       +b^{l_i}_{6,-6} {\cal O}_{6,-6}  
		                       +b^{l_i}_{6,+6} {\cal O}_{6,+6}.
\label{table2}
\end{eqnarray}

The complex coefficients $b^{l_i}_{l\pm m}$ are then linear combinations of the original
coefficients  $a_{ij}$.
As an example we list the resulting Hamiltonian obtained by rotating 
eq.\ref{H1} from $l_1$ to the laboratory system in eq.\ref{Ax}. 
All $ b^{l_1}_{l\pm m} $, which are not listed in eq.\ref{Ax} vanish for
this rotation from  $l_1$ to the laboratory system, also for this rotation no coefficients with odd j occur. 

\begin{eqnarray}
\begin{array}{|c| c|} 
 \hline 
b^{l_1}_{2,+2}   & \frac{-i}{4} (\sqrt{6} a_{20} + 2 a_{22}) \frac{\alpha_J}{f_{22}}\\

b^{l_1}_{2,0}  & \frac{1}{2}(-a_{20} + \sqrt{6} a_{22}) \frac{\alpha_J}{f_{20}}\\

b^{l_1}_{2,-2} & \frac{i}{4} (\sqrt{6} a_{20} + 2 a_{22})\frac{\alpha_J}{f_{22}}\\

 \hline 

b^{l_1}_{4,+4} &\frac{1}{16} (-\sqrt{70} a_{40} - 2 (2 \sqrt{7} a_{42} + a_{44}))\frac{\beta_J}{f_{44}}\\
b^{l_1}_{4,+2} &\frac{i}{8} (\sqrt{10} a_{40} - 4 a_{42} - 2 \sqrt{7} a_{44})\frac{\beta_J}{f_{42}}\\
b^{l_1}_{4,+0} &\frac{1}{8} (3 a_{40} - 2 \sqrt{10} a_{42} + \sqrt{70} a_{44})\frac{\beta_J}{f_{40}}\\
b^{l_1}_{4,-2}& \frac{-i}{8} (\sqrt{10} a_{40} - 4 a_{42} - 2 \sqrt{7} a_{44})\frac{\beta_J}{f_{42}}\\
b^{l_1}_{4,-4}&\frac{1}{16} (-\sqrt{70} a_{40} - 2 (2 \sqrt{7} a_{42} + a_{44}))\frac{\beta_J}{f_{44}}\\
\hline

b^{l_1}_{6,+6} & \frac{i}{32} (\sqrt{231} a_{60} + 3 \sqrt{55} a_{62} + \sqrt{66}) a_{64} + a_{66})\frac{\gamma_J}{f_{66}}\\
b^{l_1}_{6,+4} &\frac{1}{32} (3 \sqrt{14} a_{60} - \sqrt{30} a_{62} - 26 a_{64} - \sqrt{66} a_{66})\frac{\gamma_J}{f_{64}}\\
b^{l_1}_{6,+2} & \frac{-i}{32} (\sqrt{105} a_{60} - 17 a_{62} + \sqrt{30} a_{64} + 3 \sqrt{55} a_{66})\frac{\gamma_J}{f_{62}}\\
b^{l_1}_{6,0} & \frac{1}{16}(-5 a_{60} + \sqrt{105} a_{62} - 3 \sqrt{14} a_{64} + \sqrt{231} a_{66}) \frac{\gamma_J}{f_{60}}\\
b^{l_1}_{6,-2} & \frac{i}{32} (\sqrt{105} a_{60} - 17 a_{62} + \sqrt{30} a_{64} + 3 \sqrt{55} a_{66})\frac{\gamma_J}{f_{62}}\\
b^{l_1}_{6,-4} & \frac{1}{32}(3 \sqrt{14} a_{60} - \sqrt{30} a_{62} - 26 a_{64} - \sqrt{66} a_{66})\frac{\gamma_J}{f_{64}}\\
b^{l_1}_{6,-6} & \frac{-i}{32} (\sqrt{231} a_{60} + 3 \sqrt{55} a_{62} + \sqrt{66} a_{64} + a_{66})\frac{\gamma_J}{f_{66}}\\
 \hline 
\end{array}
\label{Ax}
\end{eqnarray}


\begin{references}
\bibitem{winkler}  G. Winkler, Magnetic Garnets, Vieweg, Braunschweig (1981).
\bibitem{Strohm2005} C.~Strohm et al., Phys. Rev. Lett. {\bf 96}, 155901 (2005).
\bibitem{Inyushkin2007} A.V.~Inyushin and A.N.~Taldenkov, JETP Letters {\bf 86}, 379 (2007)
\bibitem{us1} A.Sytcheva, U. L\"ow, S. Yasin, J. Wosnitza, S. Zherlitsyn, T. Goto, P. Wyder, 
and B.L\"uthi, J. Low Temp. Phys. {\bf 159}, 126 (2010).
\bibitem{us2} A. Sytcheva, U. L\"ow, S. Yasin, J. Wosnitza, S. Zherlitsyn, P. Thalmeier, T. Goto, P. Wyder, and B. L\"uthi
Phys. Rev. B {\bf 81}, 214415 (2010).
\bibitem{us3} U.~L\"ow, S.~Zvyagin, M.~Ozerov, U.~Schaufuss, V.~Kataev, B.~Wolf, and 
 B.~L\"uthi, Eur. Phys. J. B.(2013) 86:87.
\bibitem{Kama} K.~Kamazawa et al.,  Phys. Rev. B {\bf 78}, 064412 (2008).
\bibitem{araki} K.~Araki et al., Eur. Phys. J. B {\bf 61}, 257 (2008). 
\bibitem{Guillot} M.Guillot et al., J.Phys. C {\bf 18}, 3547 (1985).
\bibitem{Hutchings64} M. T. Hutchings, Solid State  Phys. {\bf 16}, 227-273 (1964).
\bibitem{Wybourne} B. G. Wybourne, Spectroscopic Properties of Rare Earths (J. Wiley and Sons, Inc., New York, 1965).
\bibitem{BL} B. L\"uthi, Physical Acoustics in the Solid State (Springer,2005), 2nd edn. 2007.
\bibitem{Levitin} R.Z.~Levitin, et. al.Physics of the Solid State, Vol. 44,
  No. 11,(2107-2111)2002.
\bibitem{Edmonds} A.R. Edmonds, Angular Momentum in Quantum Mechanics, Princeton Landmarks in Physics, (Princeton University Press 1974).
\bibitem{Hamman} J.~Hammann and P.~Manneville, J. de Phys. {\bf 34}, 615 (1973).
\bibitem{Varshny} Y.P.~Varshny, Phys. Rev. B {\bf 2}, 3952 (1970).
\bibitem{Lind} P.A. Lindg\r ard, O. Danielsen J. Phys. C {\bf 7}, 1523 (1974).
\bibitem{KLW} B.E.Keen,D.P.Landau, W.P. Wolf, Phys.Lett. {\bf 23},(1966) 202. (ver)
\bibitem{Inyushkin2010} A.V.~Inyushin and A.N.~Taldenkov, JETP Letters {\bf 111}, 760 (2010).
\bibitem{BLCL} B.L\"uthi, C.Lingner, Z. Phys. {\bf 34}, 157 (1979). 
\end{references}
\end{document}